\documentclass[twocolumn]{aastex631}
\usepackage[T1]{fontenc}
\usepackage{graphicx}
\usepackage[]{natbib}
\usepackage{amsmath, multirow}
\usepackage[colorinlistoftodos]{todonotes}

\usepackage[percent]{overpic}

\newcommand{\jwst}{\textit{JWST}}

\newcommand{\photutils}{\texttt{Photutils}}
\newcommand{\db}{\texttt{DENSE BASIS}}

\newcommand{\galfit}{\texttt{Galfit}}
\newcommand{\lenstruction}{\texttt{Lenstruction}}

\newcommand{\emcee}{\texttt{Emcee}}

\newcommand{\ba}{BulletArc-z11}
\newcommand{\zspec}{$z=11.10^{+0.11}_{-0.26}$}
\newcommand{\mulens}{$\mu=14.0^{+6.2}_{-0.3}$}
\newcommand{\lum}{$0.086^{+0.008}_{-0.030}\,L^*$}

\newcommand{\Oii}{[O~{\sc ii}]}

\newcommand{\Neiii}{[Ne~{\sc iii}]}

\usepackage{hyperref}
\begin{document}
\title{Star Formation under a Cosmic Microscope: Highly magnified $z=11$  galaxy behind the Bullet Cluster}
\shorttitle{Highly magnified $z=11$ galaxy behind the Bullet Cluster}
\shortauthors{Brada\v{c} et al.}

\author[0000-0001-5984-03925]{Maru\v{s}a Brada{\v c}}
\affiliation{University of Ljubljana, Faculty of Mathematics and Physics, Jadranska ulica 19, SI-1000 Ljubljana, Slovenia}
\affiliation{Department of Physics and Astronomy, University of California Davis, 1 Shields Avenue, Davis, CA 95616, USA}

\author[0009-0000-2101-1938]{Jon Jude\v{z}}
\affiliation{University of Ljubljana, Faculty of Mathematics and Physics, Jadranska ulica 19, SI-1000 Ljubljana, Slovenia}

\author[0000-0002-4201-7367]{Chris Willott}
\affiliation{NRC Herzberg, 5071 West Saanich Rd, Victoria, BC V9E 2E7, Canada}

\author[0009-0009-4388-898X]{Gregor Rihtar\v{s}i\v{c}}
\affiliation{University of Ljubljana, Faculty of Mathematics and Physics, Jadranska ulica 19, SI-1000 Ljubljana, Slovenia}

\author[0000-0003-3243-9969]{Nicholas S. Martis}
\affiliation{University of Ljubljana, Faculty of Mathematics and Physics, Jadranska ulica 19, SI-1000 Ljubljana, Slovenia}

\author[0000-0001-9414-6382]{Anishya Harshan}
\affiliation{University of Ljubljana, Faculty of Mathematics and Physics, Jadranska ulica 19, SI-1000 Ljubljana, Slovenia}

\author[0009-0001-0778-9038]{Giordano Felicioni}
\affiliation{University of Ljubljana, Faculty of Mathematics and Physics, Jadranska ulica 19, SI-1000 Ljubljana, Slovenia}

\author[0000-0003-3983-5438]{Yoshihisa Asada}
\affiliation{Dunlap Institute for Astronomy and Astrophysics, 50 St. George Street, Toronto, Ontario M5S 3H4, Canada}

\author[0000-0001-8325-1742]{Guillaume Desprez} 
\affiliation{Kapteyn Astronomical Institute, University of Groningen, P.O. Box 800, 9700AV Groningen, The Netherlands}

\author[0000-0003-2146-1557]{Douglas Clowe}
\affiliation{Department of Physics, Ohio University, 1 Ohio University, Athens, OH 45701, USA}

\author[0000-0002-0933-8601]{Anthony H. Gonzalez}
\affiliation{Department of Astronomy, University of Florida, Bryant Space Science Center, Gainesville, FL 32611, USA}

\author[0000-0003-2206-4243]{Christine Jones}
\affiliation{Center for Astrophysics, Harvard \& Smithsonian, 60 Garden St, Cambridge, MA 02138, USA}
\author[0000-0002-1428-7036]{Brian C. Lemaux}
\affiliation{Gemini Observatory, NSF NOIRLab, 670 N. A'ohoku Place, Hilo, Hawai'i, 96720, USA}
\affiliation{Department of Physics and Astronomy, University of California Davis, 1 Shields Avenue, Davis, CA 95616, USA}

\author[0000-0003-0144-4052]{Maxim Markevitch}
\affiliation{NASA/Goddard Space Flight Center, Greenbelt, MD 20771, USA}

\author[0000-0002-5694-6124]{Vladan Markov}
\affiliation{University of Ljubljana, Faculty of Mathematics and Physics, Jadranska ulica 19, SI-1000 Ljubljana, Slovenia}
\author[0000-0002-8530-9765]{Lamiya Mowla}
\affiliation{Whitin Observatory, Department of Physics and Astronomy, Wellesley College, 106 Central Street, Wellesley, MA 02481, USA}

\author{Gaël Noirot}
\affiliation{Department of Astronomy and Physics and Institute for Computational Astrophysics, Saint Mary's University, 923 Robie Street, Halifax, B3H 3C3, Nova Scotia}
\affiliation{Space Telescope Science Institute, 3700 San Martin Drive, Baltimore, MD 21218, USA}

\author[0000-0002-8040-6785]{Annika H. G. Peter}
\affiliation{Department of Physics, Department of Astronomy, and CCAPP, The Ohio State University}
\author[0000-0002-0086-0524]{Andrew Robertson}
\affiliation{Carnegie Observatories, 813 Santa Barbara Street, Pasadena, CA 91101, USA}
\author[0000-0001-8830-2166]{Ghassan T. E. Sarrouh}
\affiliation{Department of Physics and Astronomy, York University, 4700 Keele St., Toronto, Ontario, M3J 1P3, Canada}

\author[0000-0002-7712-7857]{Marcin Sawicki} 
\affiliation{Department of Astronomy and Physics and Institute for Computational Astrophysics, Saint Mary's University, 923 Robie Street, Halifax, B3H 3C3, Nova Scotia}

\author[0000-0002-6987-7834]{Tim Schrabback}
\affiliation{Universit\"{a}t Innsbruck, Institut f\"{u}r Astro- und Teilchenphysik, Technikerstr.
25/8, 6020 Innsbruck, Austria}
\author[0000-0002-9909-3491]{Roberta Tripodi}
\affiliation{INAF - Osservatorio Astronomico di Roma, Via Frascati 33, I-00078 Monte Porzio Catone, Italy}
\affiliation{University of Ljubljana, Faculty of Mathematics and Physics, Jadranska ulica 19, SI-1000 Ljubljana, Slovenia}
\affiliation{IFPU - Institute for Fundamental Physics of the Universe, via Beirut 2, I-34151 Trieste, Italy}

\begin{abstract}

We present measurements of stellar population properties of a newly discovered spectroscopically confirmed {\zspec}, gravitationally lensed galaxy,  using {\it JWST} NIRSpec PRISM spectroscopy and NIRCam imaging. The arc is highly magnified by the Bullet Cluster (magnification factor {\mulens}). It contains three star-forming components of which one is barely resolved and two are unresolved, giving intrinsic sizes of $\lesssim 10\mbox{pc}$.  The clumps also contain  $\sim 50\%$ of the total stellar mass. The galaxy formed the majority of its stars $\sim 150\mbox{Myr}$ ago (by $z\sim 14$). The spectrum shows a pronounced damping wing, typical for galaxies deep in the reionisation era and indicating a neutral IGM at this line of sight.  The intrinsic luminosity of the galaxy is {\lum} (with $L^*$ being the characteristic luminosity for this redshift), making it the lowest luminosity spectroscopically confirmed galaxy at $z>10$ discovered to date. 

\end{abstract}


\section{Introduction}
\label{sec:intro}

The epoch of reionisation, which signified the transformation of the universe from opaque to transparent, is poorly
understood. The cosmic microwave background (CMB) observations now
indicate that reionisation began within a few hundred
Myr of the Big Bang \citep{planck20}. However, how it proceeded and which
are the primary sources of reionisation are still the topics of much
debate. Studying the stellar properties of galaxies at the highest redshifts and a wide range of stellar mass is, therefore, necessary to make advances.  

The advent of {\jwst} imaging surveys and follow-up spectroscopy has led to a rapid advance in our knowledge of the universe at redshifts $z>10$. The space density of luminous galaxies at this epoch is greater than previous predictions \citep{harikane23a}, requiring revisions to galaxy evolution models such as enhanced early star formation \citep{dekel23}, increased burstiness \citep{sun23}, lack of dust \citep{ferrara23} and potentially a significant contribution from accreting supermassive black holes \citep{maiolino24}. Extremely luminous galaxies have been spectroscopically identified up to $z=14.4$ \citep{Carniani2024, naidu25}. It is important to measure the properties of less luminous galaxies that potentially dominate the ionising photon output of galaxies in the early Universe. 

{\jwst} with its high spatial resolution and excellent sensitivity has enabled
us to map the physics of star formation within some early
galaxies. However, NIRCam's pixel scale ($0.031\arcsec/\mbox{pix}$ at short wavelengths, equivalent to $\sim 100\mbox{pc}$ at $z=10$) and resolution do not allow the detailed studies of star formation at sub-$\mbox{100pc}$ scales. 

Already, the studies with the Hubble Space Telescope have shown that star-forming regions smaller than 100 pc play an important role in galaxies at $z\sim 3$ \citep[e.g.][]{johnson17}. However, such scales are only revealed when gravitational lensing with high magnification is present. At high redshifts $z>5$, this was recently shown by several studies. Extremely magnified
high-z galaxies have been studied in, among others, \citet{strait23,asada23,vanzella23,vanzella24,fujimoto24, bradac24, mowla24, adamo24, messa25, claeysens23, claeyssens25}. These galaxies appear to consist of several clumps, with a significant fraction of total stellar mass being in compact star-forming components. These
compact and dense structures are found with a range of stellar masses ($10^4-10^7 M_{\odot}$). They have large stellar surface densities (some with $>10^5 M_{\odot}\:\mbox{pc}^{-2}$), up to three orders of magnitude larger than what is seen in the Milky Way star clusters. Early results suggest that the majority of star formation is happening in these dense structures; however, the samples are still small.

The Bullet Cluster ($z=0.295$) is a well-known galaxy cluster merger, which was used in the past to study the properties of dark matter (\citealp{bradac06,clowe06,randall08}). Due to its merger geometry, it is also a very efficient gravitational lens \citep{bradac09}. It was recently observed by JWST (JWST-GO-4598 co-PIs Brada\v{c}, Rihtar\v{s}i\v{c}, Sawicki). In this paper, we present the highest redshift highly magnified ({\mulens}) galaxy to date. At the spectroscopic redshift of {\zspec}, we are able to study a resolved stellar population at less than $10\mbox{pc}$ scale. The galaxy, named {\ba}, is stretched into an arc with a clumpy structure. The high magnification of the galaxy allows us to increase the resolution of our observations, down to a few parsecs. 

This paper is structured as follows. In Section~\ref{sec:data} we
present the data used in this paper and in 
Section~\ref{sec:dataanalysis} we describe the analysis of the photometric and spectroscopic data. In
Section~\ref{sec:results} we present the main science results.  We summarise our results in
Section~\ref{sec:conclusions}.
Throughout the paper, we assume a
$\Lambda$CDM cosmology  with $\Omega_{\rm m}=0.3$ and Hubble constant
$H_0=70{\rm\ kms^{-1}\:\mbox{Mpc}^{-1}}$ for ease of comparison with previous work.

\section{Data}
\label{sec:data}
 {\jwst} NIRCam and NIRSpec observations of {\ba} were taken as part of the GO Program \#4598 	Silver Bullet for Dark Matter (co-PIs Brada\v{c}, Rihtar\v{s}i\v{c}, Sawicki).  The field was observed on 20 January 2025 with NIRCam imaging using filters F090W, F115W, F150W, F200W, F277W, F356W, F410M, and F444W with exposure times of $6.4 \mbox{ks}$ each, reaching S/N between 5 and 10 for a $m_{\rm AB} = 29$ point source.  We used the INTRAMODULEX 6-point dither pattern to fill in the gaps between the short-wave (SW) detectors. We also utilized archival data of {\tt HST/ACS} imaging from HST-GO-10200 (PI Jones), 10863 (PI Gonzalez), and 11099 (PI Brada\v{c}). NIRSpec Micro-Shutter Assembly (MSA) observations of targets selected from the NIRCam and ACS imaging were executed on 27 March 2025. The observations consisted of three separate MSA configurations with the low-resolution prism, using the standard 3 shutter nodding procedure with total integration time of $3.5 \mbox{ks}$ per MSA configuration. {\ba} was observed in the third MSA configuration.  
 
\begin{figure*}
      \includegraphics[width=1.0\textwidth]{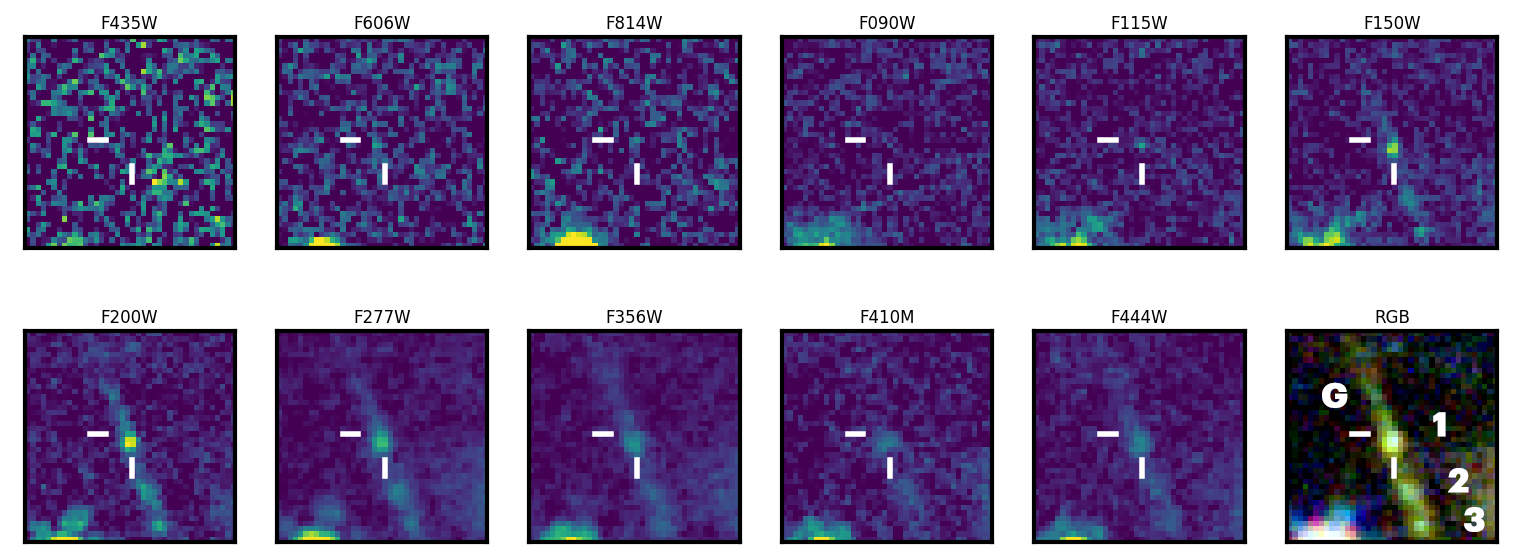}

  \centering
   \begin{overpic}[width=1.0\textwidth]{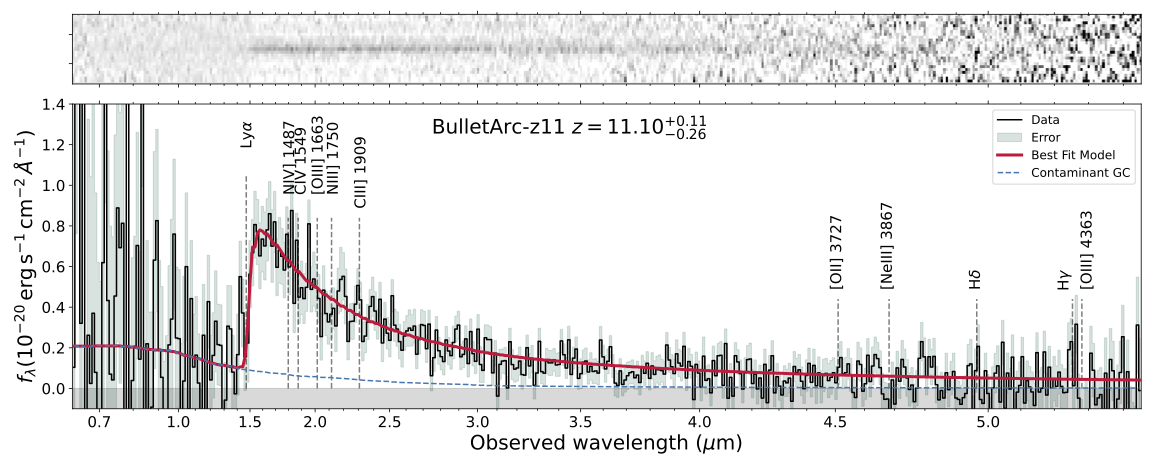}
     \put(50,18){\includegraphics[width=0.1\textwidth]{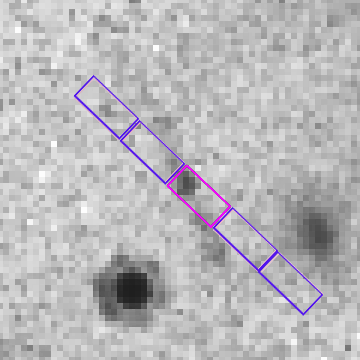}}
  \end{overpic}
  \caption{{\bf Top:} Image of the {\ba}. Shown are different filters and an RGB image (B=F090W+F115W+F150W, G=F200W+F277W+F356W, and R=F356W+F410M+F444W). The white ticks mark the central position of the source (given by R.A., Decl. in Table \ref{tab:prop}). The galaxy consists of two unresolved (1,3) and one barely resolved clump (3) and an underlying smooth component (labelled G). The sizes of cutouts are $1\farcs 6 \times 1\farcs6$. The point source near the center in the F115W filter is the contaminant globular cluster in the Bullet Cluster. {\bf Bottom:} NIRSpec prism spectrum of {\ba}. In the top panel, we show the 2D spectrum, while the bottom panel shows the 1D optimal extraction, as well as a cutout showing placement within the MSA shutters. The wavelengths of commonly observed emission lines are marked; however, none are detected with $S/N>3$. The model spectrum (red line) includes a fully neutral IGM, a neutral hydrogen damping wing column and a foreground contaminant source modeled as a globular cluster at the Bullet cluster redshift (blue dashed line). }
    \label{fig:spectrum}
 \end{figure*}

 \defcitealias{canucsdr}{Sarrouh \& Asada et al. (2025)} 
We use the CANUCS photometric pipeline presented in more detail in \citetalias{canucsdr} to reduce the imaging data. 
Bright cluster and foreground galaxies and
intracluster light are removed from the images, following the procedure outlined in \citet{martis24}. {\ba} was identified as a $z\approx11$ candidate in the NIRCam imaging with the Lyman break lying in the F150W filter. We show cutouts of the arc in Fig.~\ref{fig:spectrum}. 
 
 Details of the NIRSpec processing are given in \citetalias{canucsdr} and  \citet{heintz25}. Initial processing uses the STScI \texttt{jwst} stage 1 pipeline with custom snowball and 1/{\it f} noise correction. The \texttt{jwst} stage 2 pipeline is run up to the photometric calibration step followed by the \texttt{grizli} \citep{grizli23} and \texttt{msaexp}  \citep{msaexp} packages. The wavelength calibration uses a correction for the known intra-shutter offset along the dispersion direction. The spectral background is removed using the standard nodded background subtraction. One-dimensional spectra are extracted using a wavelength-dependent optimal extraction that accounts for the increase in PSF FWHM with wavelength.

\begin{figure*}
\includegraphics[width=1.0\textwidth]{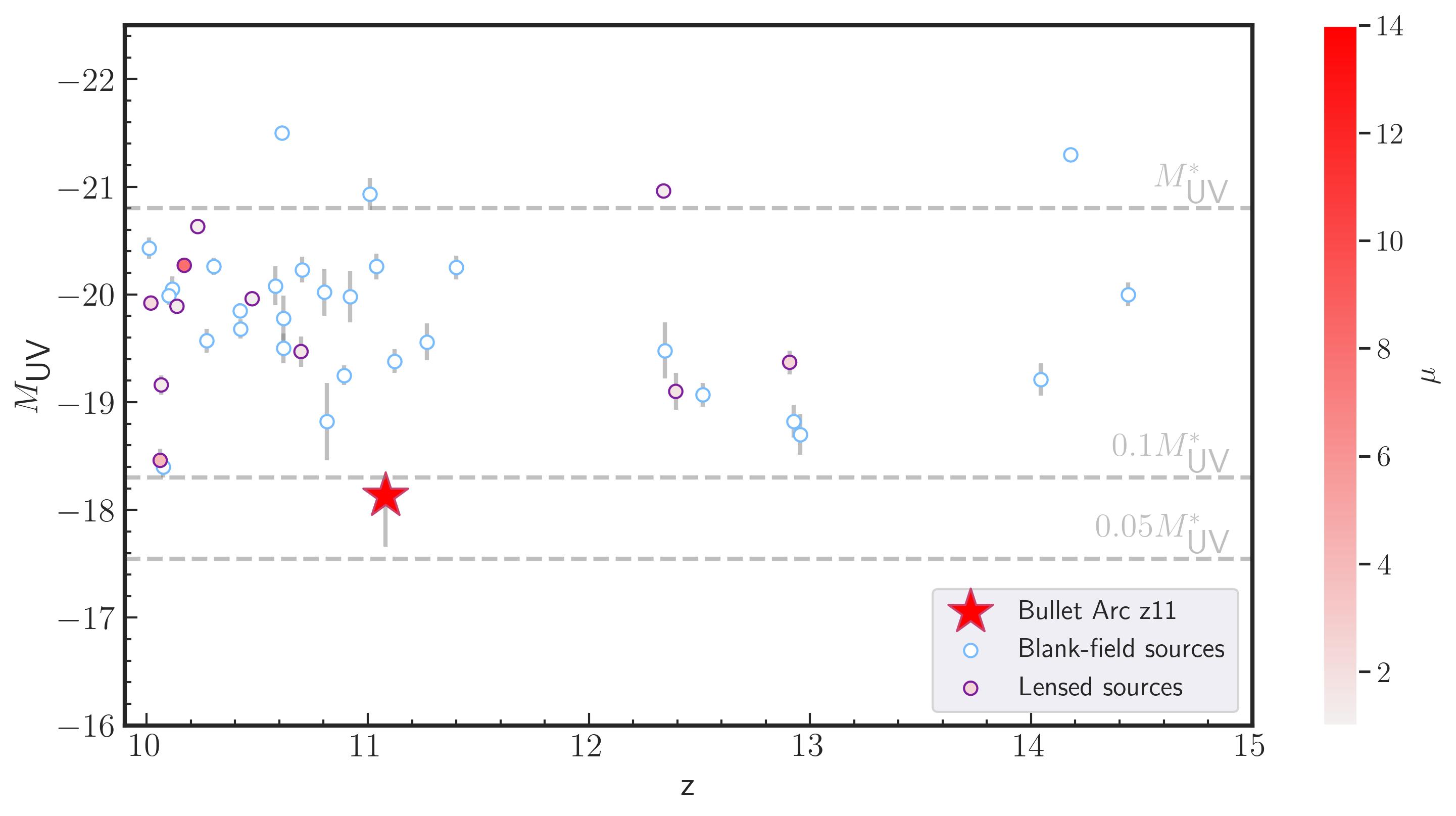}
  \caption{$M_{\rm UV}$ vs redshift plot for the {\ba} (star) and spectroscopically confirmed sources from the literature (circles, collated in \citealp{robertsborsani25}). Fill colours represent magnification $\mu$ given by the colorbar. Open circles are blank-field sources with $\mu=1$. $M_{\rm UV}$ values for different characteristic luminosities $L^*$ (at $z\sim 10$, taken from \citealp{willott24}) are shown as dashed horizontal lines. The uncertainty on $M_{\rm UV}$ for {\ba} is partly hidden behind the symbol and includes lensing uncertainties.}
    \label{fig:Muvz}
 \end{figure*}

\section{Data Analysis}
\label{sec:dataanalysis}

\subsection{Spectroscopy}\label{sec:spec}
The central bright clump of {\ba} (\#1 in Fig.\ref{fig:spectrum}) was the target position for the NIRSpec MSA observation. The resulting spectrum and the slit placement are shown in Fig.~\ref{fig:spectrum}. No emission lines are securely detected with $S/N>3$. However, a clear Lyman break is observed at 1.47\,$\mu$m, confirming the high-redshift nature of this galaxy. The smooth rollover at the Lyman break is suggestive of strong Lyman-$\alpha$ absorption from a neutral IGM, potentially with additional absorption from a proximate dense neutral absorber.

The redshift is determined by fitting a power-law continuum model with the Monte-Carlo Markov-Chain code {\emcee} \citep{Foreman-Mackey2013}. The model contains four free parameters, corresponding to the normalization, redshift, $\beta_{\rm UV}$ slope, and neutral hydrogen damping wing column, $N_{\rm HI}$. The neutral hydrogen damping wing was modeled as in \cite{deugenio24} and a fully neutral intergalactic medium (IGM) was assumed. 

The point source contaminant described in the following subsection can be seen in the spectrum as a faint flux excess below the Lyman break. The F090W-F115W color and compactness are fully consistent with other numerous globular clusters (GCs) in the field of the Bullet Cluster. We thus model the contaminant as a GC at a redshift $z=0.295$  with a simple stellar population with metallicity 0.1 solar and age at redshift 0 of 12\, Gyr (c.f. \citealt{harris25}). We use E-MILES template\footnote{\url{https://research.iac.es/proyecto/miles/pages/spectral-energy-distributions-seds/e-miles.php}},  normalized by the observed spectrum flux between 0.85\,$\mu$m and 1.4\,$\mu$m where the flux from {\ba} should be zero. The resulting model is shown in Fig.~\ref{fig:spectrum}. This `GC contaminant' model is subtracted from the spectrum prior to fitting. 

The resulting redshift is {\zspec} (Table~\ref{tab:prop}), with the significant uncertainty due to the degeneracy with $N_{\rm HI}$. The value of $N_{\rm HI}$ is not meaningfully constrained in the absence of an emission line redshift. Despite the latter, the high redshift solution is robust. Using full photometry (Sect.~\ref{sec:phot}) the redshift of the source is $z_{\rm phot} = 11.1 ^{+0.9}_{-1.3}$ and there are no solutions below redshift 6.

From the spectral fit, we also find $\beta_{\rm UV} = -2.44 ^{+0.17}_{-0.18}$. Because of possible uncertainty in the wavelength-dependent slit losses, we adjust the spectrum with the NIRCam photometry (using the custom photometric aperture with $S/N=16$, see next section) following \citet[in prep.]{felicioni25}, to obtain $\beta_{\rm UV} =-2.3\pm 0.2$ (Table~\ref{tab:prop}).

\subsection{Photometry}\label{sec:phot}
Photometry is derived using the updated
zero-points, and corrected for Milky Way extinction using the value of colour excess $E(B-V)=0.207$ from \citealp{schlafly11}. We also assume the extinction law by \citet{fitzpatrick99} using the factor between the extinction coefficient and colour excess $R_{\rm V}=3.1$. {\ba} is detected in the F150W,
F200W, F277W, F356W, F410M, and F444W NIRCam filters (Fig.~\ref{fig:spectrum}). In the F115W filter, there is a compact source detected just 40 milliarcsec north-west of the brightest component of {\ba}, which we also see in the spectrum (Sect~\ref{sec:spec}). As noted, this source has colors and compactness consistent with a globular cluster within the intracluster region of the Bullet cluster. There are many similar sources in this region of the NIRCam imaging.   We do not subtract the flux of the contaminant GC, since it is negligible, as evidenced by the spectrum and measured F115W flux of 2\,nJy. 

For the photometry of the entire arc, we initially follow the CANUCS pipeline and procedure as described in \citetalias{canucsdr}. We build a $\chi_{\rm mean}$-detection image \citep{Drlica-Wagner2018a} by combining all images before any PSF convolution. Due to high distortion of the arc, standard photometric apertures are not optimal. To measure the total stellar mass and luminosity we use a custom aperture including all pixels above a detection image threshold of $S/N=3$. To measure the most accurate colors we use the custom aperture with the highest average $S/N$ across all filters at $\geq 2\mu$m, which corresponds to pixels with $S/N>16$ in the detection image. All fluxes are measured on images that have been convolved to the PSF of the F444W image to ensure the same regions contribute at all wavelengths. For the total flux we additionally scale the PSF-convolved fluxes by the ratio of the non-convolved to convolved F200W fluxes to take account of flux scattered out of the custom aperture during the PSF convolution process. Flux uncertainties are determined from the distribution of sky values measured in identically-shaped apertures located in empty regions of the image.

We present fluxes in Table~\ref{tab:phot}, using label total for the total flux and SN16 for the optimal S/N aperture for the best colours. The source has total F200W AB magnitude of $26.63\pm 0.07$, giving absolute magnitude (corrected for lensing magnification and uncertainties) of $M_{\rm UV}= -18.13^{+0.47}_{-0.09}$. Using the characteristic luminosity $L^*$ from \citet{willott24} results in a luminosity of {\ba} {\lum}. We find {\ba} is the lowest luminosity spectroscopically confirmed galaxy at $z>10$ discovered to date (Fig.~\ref{fig:Muvz}). The photometry shows a marked upturn at $>4.3\,\mu$m, similar to but stronger than that observed in GN-z11 \citep{tacchella23a}. Corresponding to rest-frame wavelengths $>3600\,$\AA, this could be due to a Balmer break and/or strong higher-order Balmer, \Oii\ and \Neiii\ emission lines. Such emission lines are not ruled out by the non-detections in our shallow prism spectroscopy.

Since {\ba} is resolved into three distinct clumps and a smooth galaxy component, we also perform spatially resolved photometry. We fit the four components using three point sources (for the clumps, below we also perform forward modelling with extended components to measure sizes) and a Gaussian profile (for the underlying galaxy distribution G) in the F200W image with {\galfit} \citep{peng11}. Their central coordinates were then used as fixed priors to repeat the process in all filters and extract the photometry of the four components. Errors are measured by injecting artificial sources of similar brightness. This photometry is also given in Table~\ref{tab:phot}.

\begin{figure}
\includegraphics[width=0.5\textwidth]{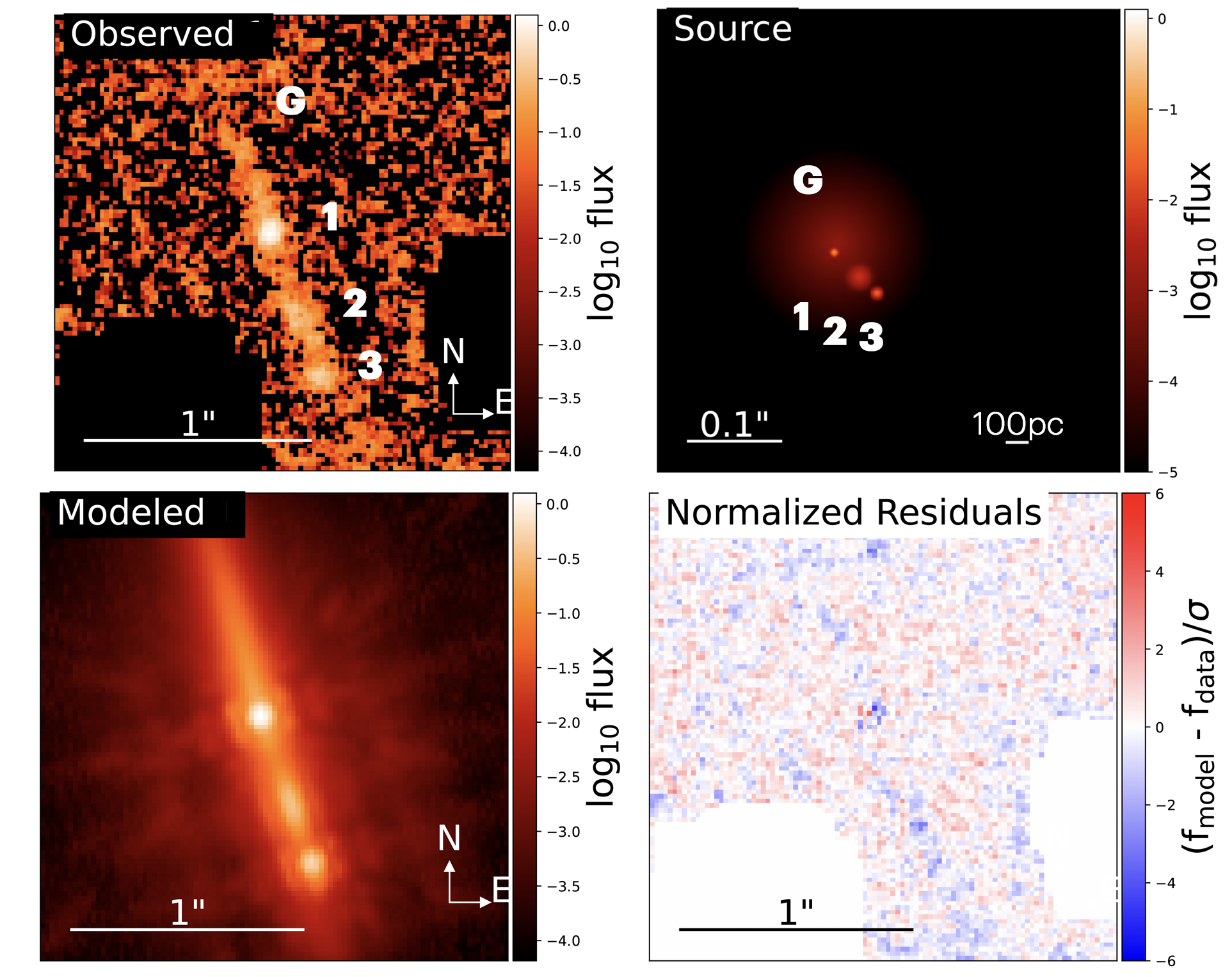}
  \caption{Source plane modelling using {\lenstruction} \citep{yang20}. From the top left, we plot the observed image (in F200W),  the source model, the best fit image model (after lensing and PSF convolution have been applied), and normalised residuals. The source was modelled using an underlying galaxy component G and 3 small clumps (\#1-3) modelled using Gaussian profiles.}
    \label{fig:lenstruction}
 \end{figure}

Because of high lensing distortion, we also measure sizes and positions in the source plane by forward modelling the observed photometry with {\lenstruction} \citep{yang20}.  We use a four-component model consisting of a Sersic profile for the underlying galaxy and Gaussian profiles for the three clumps. The source model is then distorted, using gravitational lensing maps derived by \citet[][in prep.]{rihtarsic25b} (see also Sect.~\ref{sec:lm}) and convolved with the empirical point spread function (PSF), determined from the observations of stars. The resulting image is then compared to the observations to obtain the best possible size measurement parameters.  The results are shown in Fig.~\ref{fig:lenstruction}.

\subsection{Gravitational lens model} \label{sec:lm}
\begin{figure}
\includegraphics[width=0.5\textwidth]{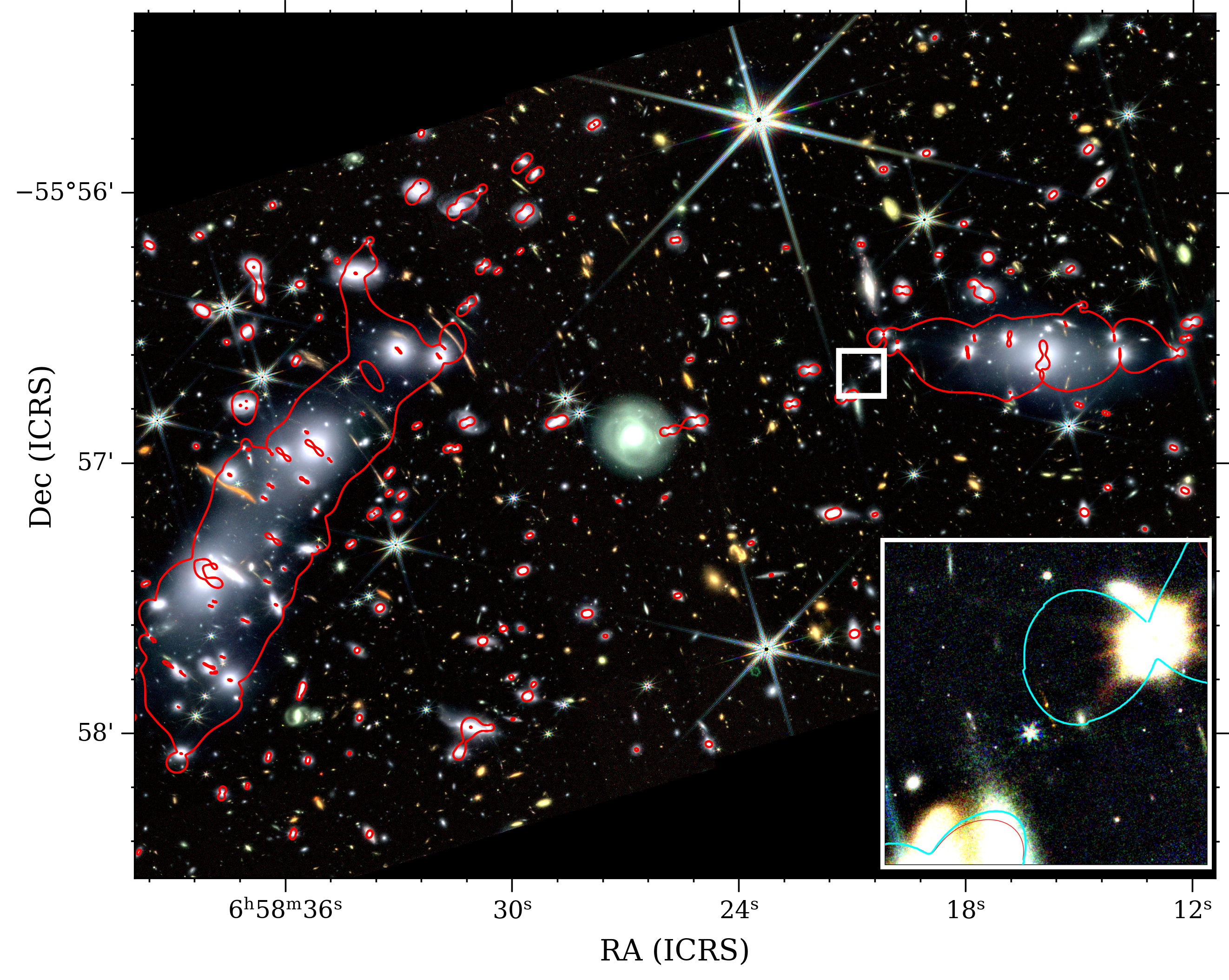}
  \caption{RGB (F277W, F357W, F444W) image of the Bullet Cluster. Overlaid (in red) is the critical curve calculated at the redshift of the {\ba} used in this paper. In the inset ($10\arcsec \times 10 \arcsec$) we use high-resolution RGB images (F115W, F150W, F200W) to show the {\ba} and the same red critical curve as well as in cyan the one with a cluster member added as described in Sect.~\ref{sec:lm}.}
    \label{fig:ccurve}
 \end{figure}

A strong lensing model was derived using the parametric lens modelling tool \texttt{Lenstool} \citep{Kneib96,Jullo07,Jullo09} and is constrained using a catalogue of 141 multiple images from 52 multiply lensed systems (28 distinct galaxies) with spectroscopic redshifts, mostly obtained from our JWST program. Prior models have used limited number of spectroscopic redshifts available for multiply imaged sources ($6$, \citealp{richard21, cha25}). The lens model will be published in \citet[in prep.]{rihtarsic25b} and includes large-scale dark matter halos, group-scale substructure halos surrounding the inner cluster regions, cluster gas, included as a mass map from Chandra X-ray observations, and cluster member halos with parameters, scaling with NIRCam F277W magnitude. {\ba} is situated in a highly magnified region, $40''$ from the subcluster BCG in a region with high gas density (Fig.~\ref{fig:ccurve}). It is furthermore highly influenced by a magnitude 18.97 cluster member $\sim4''$ away (seen in the upper-right corner of the inset in Fig.~\ref{fig:ccurve}). When modeled with scaling relations, the cluster member mass is likely overestimated, as it predicts additional {\ba} counter-images nearby for which we find no strong evidence in NIRCam imaging. If the two fainter clumps are multiply imaged, the position of the critical curve demands the brighter clump to have a counterimage.  Depending on the true mass of the cluster member, it can produce arbitrarily high magnifications of {\ba}. In the absence of strong evidence for additional multiple images that would constrain its mass individually, we derived magnifications with the cluster member excluded. We note that our results conservatively represent lower limits on magnification, and hence upper limits on derived sizes and fluxes. The {\ba} magnification, derived from this model, is {\mulens}, with the lowest limit of $\mu>10.4$, derived from the Bayesian samples of the lens model and is used in subsequent analysis. We also note that magnification values across the arc do not vary significantly (less than $\sim10\%$, which is within the error bars). For the subset of models with the above-mentioned cluster member added, that do not produce multiple images for the {\ba}, the magnification is $\mu_{\rm CM}=130^{+90}_{-70}$. Such a magnification would decrease stellar masses and star formation rates by an order of magnitude. Similarly, the tangential magnification increases by a factor of $\sim 10$, hence also the sizes would decrease by an order of magnitude.

\section{Results}
\label{sec:results}

We performed the SED fitting to the photometry (spectrum S/N is too low for a more detailed analysis) using the {\db}
method \citep{iyer17, iyer19} to determine
nonparametric star formation histories (SFHs), masses, and ages for our 
sources in {\ba}. We adopt the
Calzetti attenuation law \citep{calzetti01}, flat dust prior and a Chabrier IMF \citep{chabrier03}. We fix the redshift to the spectroscopic redshift, allowing it to vary within the uncertainties given in Table~\ref{tab:prop} (we also generate the atlas using these values).  All other parameters are left free. The primary advantage of using {\db} is that it allows us
to account for flexible star formation histories.

\begin{figure}
\centerline{\includegraphics[width=0.5\textwidth]{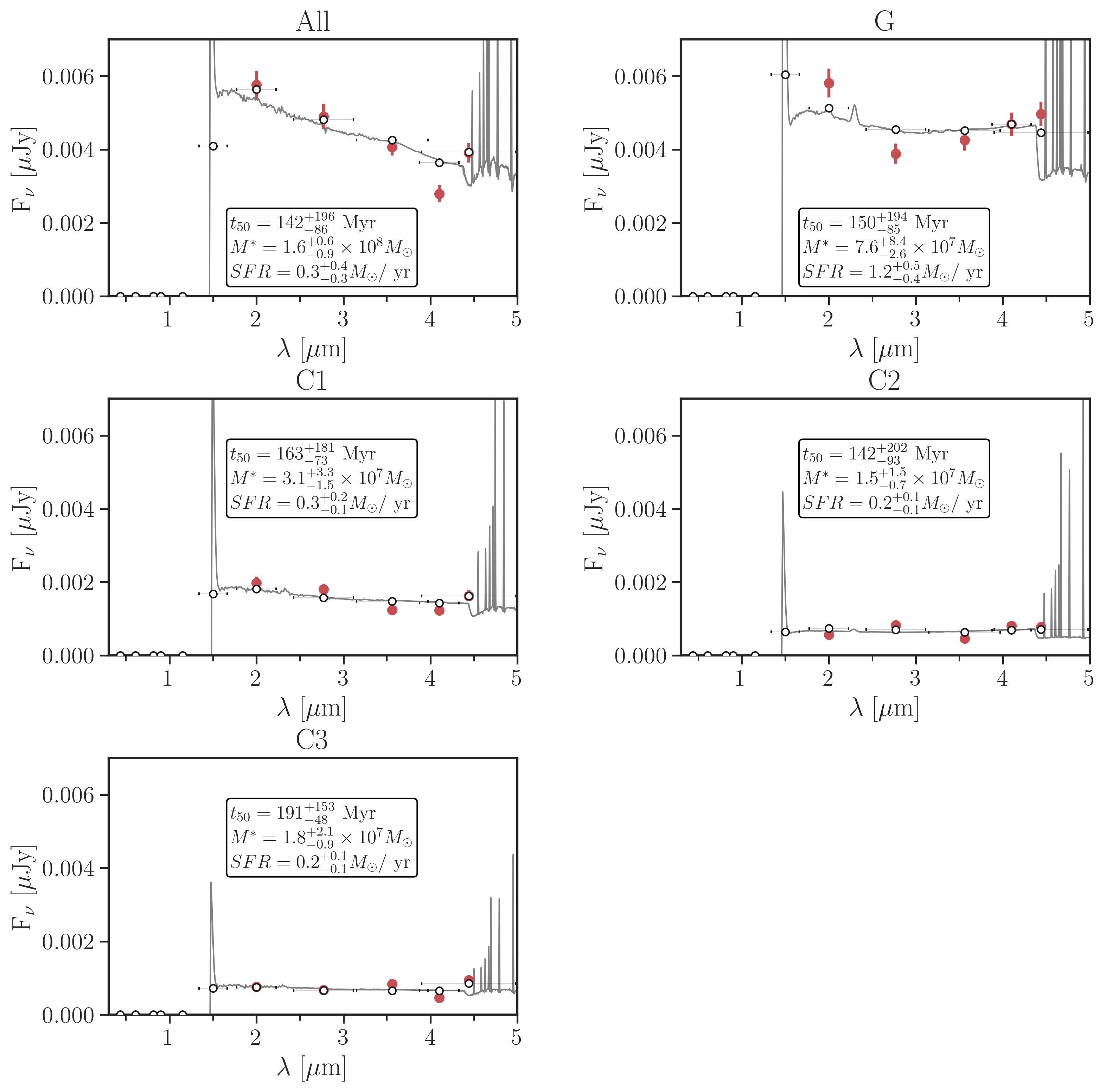}}
 \caption{Results of the SED fitting using original full photometry using {\photutils} corrected by dividing by $\mu$ (labelled All, see also Table~\ref{tab:phot}), and derived using {\galfit} for the smooth light component (G) and the three clumps (labelled C1-3). Shown are fluxes in red and SED predicted fluxes in open circles in units of $\mu\mbox{Jy}$. Derived stellar properties are given in the inset. While fluxes in F150W and shortward are shown, they were not included in the fit as the redshift is only allowed to vary based on spectroscopic redshift measurement.}
    \label{fig:SED}
\end{figure}

{\ba} has intrinsic (corrected for magnification) total stellar mass of $1.6^{+0.6}_{-0.9} \times 10^8 M_{\odot}$ and star formation rate (SFR) of  $0.29^{+0.42}_{-0.28}M_{\odot}\mbox{yr}^{-1}$ (Fig.~\ref{fig:SED}). The galaxy formed 50\% of its total mass by $t_{50}=140^{+ 200}_{-90} \mbox{Myr}$. At face value, this is relatively old, given that the age of the universe is just 400 Myrs old at this redshift, meaning that {\ba} might have formed 50\% of its stellar mass by redshift $z\sim 14$. This is similar mass to the $z=14.4$ galaxy with a stellar mass of $1.2^{+1.3}_{-0.4}\times10^8 M_{\odot}$ \citep{naidu25}. However, also note that the uncertainties are large. The measurements are also prone to systematics, given that we only have access to rest-frame UV colours of this object.

\begin{figure}
\centerline{\includegraphics[width=0.5\textwidth]{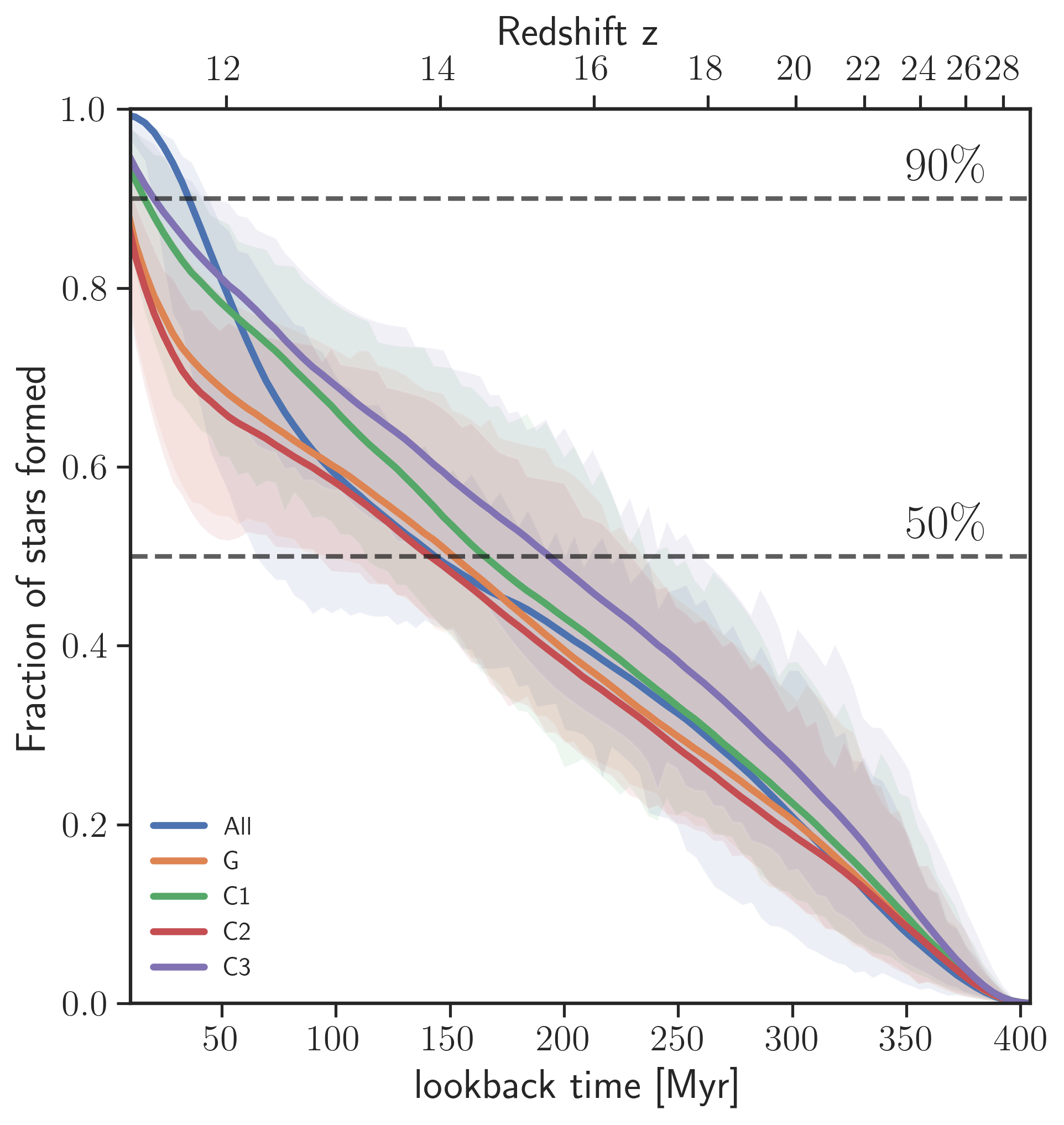}}
 \caption{Mass fraction of stars formed as a function of look-back time (bottom axis) and redshift (top axis) for all components (All-blue, galaxy-orange, C1-green, C2-red, C3-violet). All components seem to be forming 50\% of their total mass by the redshift of $\sim 14$ ($t_{50}\sim 150{\rm Myr}$). Note, however, that the results indicate a large range of possible SFHs (as indicated by the errors given as the shaded regions).}
    \label{fig:fraction}
\end{figure}

The SED fitting results for the individual components are also listed in Table~\ref{tab:prop} and shown in Fig.~\ref{fig:SED}. Note that the galaxy component as determined by {\galfit} includes the total light (beyond the aperture of the detected source). Hence, the stellar mass of all components is larger than what we measure as the total stellar mass. The clumps themselves contain $\sim 50\%$ of the measured stellar mass of the system. All components show similar star formation histories (Fig.~\ref{fig:fraction}).

Using the SW F200W band image (with highest S/N and smallest PSF), we also measure sizes of each of the components. We model the source in the source plane using {\lenstruction} with a combination of an underlying galaxy component with a round Sersic profile and a fixed Sersic index of 1. Its half-light radius is $144^{+18}_{-15}~\mbox{pc}$, consistent with size-mass relation found by \citet{morishita24}. The other three components are barely resolved (clump 2) or unresolved (clumps 1 and 3) and are considered to be a Gaussian (which produces a better fit than Sersic profiles). All the sizes are reported as the half-light radii and listed in Table~\ref{tab:prop}. 

The clumps have very small sizes, similar to those reported in other high-z gravitationally lensed sources, e.g., \citet{mowla24,adamo24}. The stellar surface mass density of clumps 1 and 3 is $>10^5M_{\odot}/\mbox{pc}^2$, which is three orders of magnitude higher than typical young star clusters in the local universe. Clump 2 also has a high stellar surface mass density of $\sim 2\times 10^4M_{\odot}/\mbox{pc}^2$. Given stellar mass and proximity to the main galaxy's centre of clump 1 (Table~\ref{tab:prop}), clump 1 could also be an AGN, although we do not see any other signatures. We investigated the possible presence of X-ray emission in very deep (0.5Ms) Chandra data. Unfortunately, {\ba} is located in the region of the bullet gas itself; therefore, the sensitivity is low. The 3-sigma upper limit for a 
point-source flux in the $0.5\text{--}2\mbox{~keV}$ band (observer frame) is $<4.9\times 10^{-7} \mbox{phot/s/cm}^2$. Given the redshift and magnification of the source, this limit corresponds to an
intrinsic luminosity in the corresponding rest-frame $6\text{--}24\mbox{~keV}$ band of
$<1.1\times 10^{44} \mbox{erg/s}$.

\section{Conclusions} \label{sec:conclusions}
We report a new discovery of a {\zspec}, highly magnified gravitationally lensed galaxy behind the Bullet cluster. The NIRSpec spectrum of {\ba} shows a distinct damping wing, indicating the presence of a neutral IGM, typical for a galaxy deep in the reionization era.  This is the lowest luminosity ({\lum}) $z>10$ galaxy spectroscopically confirmed to date. Its high magnification ({\mulens}) not only allows for its detection, but it also gives an excellent opportunity to study star formation at ten parsec scales at high redshift.

The galaxy consists of the main component and three barely/unresolved clumps. The clumps show a very high stellar surface mass density ($>10^5M_{\odot}/\mbox{pc}^2$ for clumps 1 and 3 and $\sim 2\times 10^4M_{\odot}/\mbox{pc}^2$ for clump 2). This is similar to other studies of highly magnified objects, where we see star formation broken down into smaller clumps when gravitational lensing allows us to observe at high enough resolutions. 

The central brightest clumps could either indicate the presence of an AGN or a very compact star cluster. We would require deeper/higher resolution NIRSpec spectroscopy to establish the presence of possible broad and high ionisation lines to test this hypothesis.

\section*{Acknowledgements}
MB, JJ, GR, NM, AH, GF, and VM acknowledge support from the ERC Grant FIRSTLIGHT, Slovenian national research agency ARIS through grants N1-0238 and P1-0188, and ESA PRODEX Experiment Arrangement No. 4000146646. AP, AR acknowledge support from grant JWST-GO-4598. This research was also enabled by grant 18JWST-GTO1 from the Canadian Space Agency and funding from the Natural Sciences and Engineering Research Council of Canada. This research used the Canadian Advanced Network For Astronomy Research (CANFAR) operated in partnership by the Canadian Astronomy Data Centre and The Digital Research Alliance of Canada with support from the National Research Council of Canada the Canadian Space Agency, CANARIE and the Canadian Foundation for Innovation. The data were obtained from the Mikulski Archive for Space Telescopes at the Space Telescope Science Institute, which is operated by the Association of Universities for Research in Astronomy, Inc., under NASA contract NAS 5-03127 for JWST. These observations are associated with program \#4598.  Support for program \#4598 was provided by NASA through a grant from the Space Telescope Science Institute, which is operated by the Association of Universities for Research in Astronomy, Inc., under NASA contract NAS 5-03127. BL is supported by the international Gemini Observatory, a program of NSF NOIRLab, which is managed by the Association of Universities for Research in Astronomy (AURA) under a cooperative agreement with the U.S. National Science Foundation, on behalf of the Gemini partnership of Argentina, Brazil, Canada, Chile, the Republic of Korea, and the United States of America.

\section*{Data Availability}
All the {\it JWST} data used in this paper can be found in MAST: \dataset[10.17909/aks9-8p38]{http://dx.doi.org/10.17909/aks9-8p38}.

\facilities{HST (ACS,WFC3), JWST (NIRCam, NIRSpec)}

\bibliographystyle{aasjournal}
\bibliography{bibliogr_cv,bibliogr_highz, canucs_dr1} 

\appendix

\section{Photometry and SED fitting}
\label{sec:app}
In Table~\ref{tab:phot} we list photometry and in Table~\ref{tab:prop} derived quantities and results of SED fitting of {\ba}. All the procedures are described in the main text. 
\begin{deluxetable*}{lccccc} 
\tabletypesize{\footnotesize}
\tablecolumns{5}
\tablewidth{0pt}
\tablecaption{Properties of {\ba} and its components\label{tab:prop}.}
\tablehead{\colhead{Parameter} & \colhead{}& \colhead{}& \colhead{}& \colhead{} \\ 
\colhead{} & \colhead{All} & \colhead{G} &  \colhead{C \#1}& \colhead{C \#2} & \colhead{C \#3}}
\startdata
R.A. (deg) &  104.5865130 & & & &\\
Decl. (deg) & -$55.94450726$& & & & \\
$z_{\rm spec}$ &  {\zspec}& & & & \\
$\mu$\tablenotemark{a}   &{\mulens} & &&  & \\
$\beta_{\rm UV}$  &$-2.3\pm 0.2$ & &&  & \\
$M^* (M_{\odot})$\tablenotemark{b}& $1.6^{+0.6}_{-0.9} \times 10^8 $ & $7.6^{+8.4}_{-2.6} \times 10^7 $& $3.1^{+3.3}_{- 1.5} \times 10^7$ & $1.5^{+1.5}_{- 0.7} \times 10^7$ & $1.8^{+2.1}_{- 0.9} \times 10^7$\\
$SFR (M_{\odot}/\mathrm{~Myr})$\tablenotemark{b} &  $0.29^{+ 0.42}_{- 0.28}$ & $1.20^{+ 0.51}_{- 0.42}$ & $0.29^{+ 0.16}_{- 0.15}$ & $0.20^{+ 0.15}_{- 0.09}$ & $0.15^{+ 0.13}_{- 0.04}$\\
$t_{50} \mathrm{~Myr}$ & $143^{+ 197}_{-87}$ & $151^{+ 194}_{-85}$ & $163^{+ 182}_{-73}$ & $143^{+ 202}_{-94}$ & $192^{+ 154}_{-48}$\\
\hline
Sizes (arcsec)\tablenotemark{b,c} & &$0.037^{+0.005}_{-0.004}$ & $<0.0025$ &  $0.0062^{+0.0011}_{-0.0023}$ & $< 0.0028$\\
Sizes ($\mathrm{parsec}$)\tablenotemark{b,c}& &$144^{+18}_{-15}$ & $<10$ &  $24^{+4}_{-9}$ & $< 11$\\
Position (parsec)\tablenotemark{b} & & $(0,0)$ & $(-11,-39)$ & $(94,-140)$ & $(168,-208)$\\
2D Distance (parsec)\tablenotemark{b,d} & & 0 & $40\pm 20$ & $169\pm 20$ & $267\pm 20$\\
\enddata
\tablenotetext{a}{Magnification values across the arc do not vary significantly (see text).}
\tablenotetext{b}{All relevant quantities are determined in the source plane/have been delensed.}
\tablenotetext{c}{Quoted are half-light radii.}
\tablenotetext{d}{2D distances in the source plane from the central galaxy component(G).}
\end{deluxetable*}

\begin{deluxetable}{lcccccc} 

\tabletypesize{\footnotesize}
\tablecolumns{5}
\tablewidth{0pt}
\tablecaption{{\jwst} photometry of {\ba}. Fluxes are given for the total, SN16 (the custom aperture maximizing the $S/N$) and all four fitted components - G,C\#1-3 of {\ba}.\label{tab:phot}}
\tablehead{\colhead{Filter} & \colhead{Total} & \colhead{SN16} & \colhead{G} &\colhead{C \#1}& \colhead{C \#2} & \colhead{C \#3} \vspace{-0.2cm} \\
\colhead{} & \colhead{(nJy)}& \colhead{(nJy)}& \colhead{(nJy)}& \colhead{(nJy)} & \colhead{(nJy)}& \colhead{(nJy)}}
\startdata
F090W & $-2 \pm 5$ & $ -0.1 \pm 0.7 $ & -- & -- & -- & -- \\
F115W & $ 2 \pm 5$ & $  1.5 \pm 0.6 $ & -- & -- & -- & --\\
F150W & $47 \pm 5$ & $  9.6 \pm 0.7 $ & -- & -- & -- & -- \\
F200W & $81 \pm 5$& $16.0 \pm 0.7$ & $81 \pm 6$ & $28 \pm 2$ & $8 \pm 1$ & $11 \pm 2$ \\
F277W & $69 \pm 5$& $13.9 \pm 0.6$ & $54 \pm 4$ & $25 \pm 2$ & $12 \pm 2$ & $9 \pm 1$ \\
F356W & $57 \pm 3$& $10.6 \pm 0.5$ & $60 \pm 4$ & $17 \pm 2$ & $6 \pm 1$ & $12 \pm 2$ \\
F410M & $39 \pm 3$& $7.8 \pm 0.7$ & $66 \pm 4$ & $17 \pm 2$ & $11 \pm 2$ & $7 \pm 1$ \\
F444W & $55 \pm 4$& $10.6 \pm 0.7$ & $70 \pm 5$ & $23 \pm 2$ & $11 \pm 2$ & $13 \pm 2$ \\
\enddata
\end{deluxetable}

\end{document}